\documentclass[runningheads]{llncs}
\usepackage{cite}
\usepackage{graphicx}
\usepackage{xspace}
\usepackage{url}
\usepackage{acronym}
\usepackage{CJKutf8}
\usepackage{array}
\usepackage{xcolor,colortbl}
\usepackage{adjustbox}
\usepackage{latexsym}
\usepackage{courier}
\usepackage{subfig}
\usepackage{tabularx}
\usepackage{xcolor}
\usepackage{float}
\usepackage{algpseudocode}
\usepackage{comment}
\usepackage{booktabs}
\usepackage{longtable}
\usepackage{cancel}
\usepackage[misc]{ifsym}
\usepackage{orcidlink}

\usepackage{amssymb}
\usepackage{array}
\usepackage{algorithm}
\usepackage{algpseudocode}
\usepackage{amsmath}
\definecolor{added}{RGB}{0,150,0}
\definecolor{neutral}{RGB}{0,0,255}
\definecolor{rephrased}{RGB}{255,0,0}

\usepackage{multirow}
\begin{document}
\title{Talking Like a Phisher: LLM-Based Attacks on Voice Phishing Classifiers}
\titlerunning{Talking Like a Phisher - Accepted by ICDF2C 2025}
%
\author{WENHAO LI\inst{1}~\orcidlink{0009-0007-4342-6676},
SELVAKUMAR MANICKAM\inst{1}\textsuperscript{\Letter}~\orcidlink{0000-0003-4378-1954},
YUNG-WEY CHONG\inst{2}~\orcidlink{0000-0003-1750-7441} and
SHANKAR KARUPPAYAH\inst{1}~\orcidlink{0000-0003-4801-6370}}
\authorrunning{W. Li et al.}
%

\institute{Cybersecurity Research Centre, Universiti Sains Malaysia, Pulau Pinang, Malaysia \and
School of Computer Sciences, Universiti Sains Malaysia, Pulau Pinang, Malaysia\\
\email{wenhaoli@ieee.org, \{selva, chong, kshankar\}@usm.my}}
\maketitle              
\begin{abstract}
Voice phishing (vishing) remains a persistent threat in cybersecurity, exploiting human trust through persuasive speech. While machine learning (ML)-based classifiers have shown promise in detecting malicious call transcripts, they remain vulnerable to adversarial manipulations that preserve semantic content. In this study, we explore a novel attack vector where large language models (LLMs) are leveraged to generate adversarial vishing transcripts that evade detection while maintaining deceptive intent. We construct a systematic attack pipeline that employs prompt engineering and semantic obfuscation to transform real-world vishing scripts using four commercial LLMs. The generated transcripts are evaluated against multiple ML classifiers trained on a real-world Korean vishing dataset (KorCCViD) with statistical testing. Our experiments reveal that LLM-generated transcripts are both practically and statistically effective against ML-based classifiers. In particular, transcripts crafted by GPT-4o significantly reduce classifier accuracy (by up to 30.96\%) while maintaining high semantic similarity, as measured by BERTScore. Moreover, these attacks are both time-efficient and cost-effective, with average generation times under 9 seconds and negligible financial cost per query. The results underscore the pressing need for more resilient vishing detection frameworks and highlight the imperative for LLM providers to enforce stronger safeguards against prompt misuse in adversarial social engineering contexts.

\keywords{Adversarial Attacks \and Cybercrime \and Large Language Models (LLMs) \and Voice Phishing \and Phishing Detection.}
\end{abstract}
\section{Introduction}

Phishing is a form of cybercrime in which adversaries deceive users into disclosing sensitive information by impersonating trustworthy entities \cite{pujara2018phishing}. Despite numerous detection mechanisms being proposed, attackers continuously devise novel methods to evade them \cite{10798104,tiis:101917}. As phishing continues to evolve, it poses significant threats to individuals, organizations, and global cybersecurity, leading to substantial financial and data losses \cite{Tian04052025}.

Voice phishing (vishing) is a type of phishing attack where scammers (vishers) use phone calls to impersonate trusted organizations and trick victims into revealing sensitive information or transferring money \cite{cho2012voice,choi2017voice}. These attacks typically involve scripted conversations that exploit urgency or fear, using pretexts like tax refunds, legal threats, or delivery issues \cite{rayexploring}.

To combat these threats, researchers have developed machine learning (ML) and natural language processing (NLP)-based detection systems that analyze transcribed vishing calls for malicious patterns \cite{10901962}. However, these models remain vulnerable to subtle linguistic manipulations that preserve semantic intent while evading classification \cite{10.1145/3593042}. Recent advances in large language models (LLMs) offer new possibilities for crafting such adversarial inputs \cite{Gallagher2024}, yet their ability to generate evasive vishing transcripts remains underexplored.

To address this gap, this study proposes a systematic approach to investigate LLM-assisted adversarial vishing attacks. By prompting commercial LLMs with original scam transcripts, we generate linguistically obfuscated versions and evaluate their ability to bypass trained ML-based vishing detectors while preserving the semantic meaning. The major contributions of this study are as follows:
\begin{itemize}
    \item We propose a threat model and an LLM-assisted vishing attack pipeline that combines prompt engineering with semantic obfuscation techniques.
    \item We evaluate the effectiveness of adversarial transcripts against multiple ML-based classifiers trained on a real-world Korean vishing dataset (KorCCViD) with statistical testing.
    \item We assess semantic consistency using BERTScore to ensure the preservation of malicious intent in generated transcripts.
    \item We provide a case study on LLM-generated adversarial transcripts, analyze the practical and security implications of using commercial LLMs in adversarial vishing settings.
\end{itemize}

The structure of the rest of the paper is as follows: Section~\ref{sec:related} reviews related works. Section~\ref{sec:threat_model} outlines the threat model. Section~\ref{sec:methodology} presents our proposed methodology. Section~\ref{sec:exp} details the experimental setup. Section~\ref{sec:results} discusses evaluation results. Section~\ref{sec:con} concludes the paper.

\section{Related Work}\label{sec:related}

This section reviews related works on ML-based voice phishing detection, adversarial attacks in NLP, and the emerging role of LLMs in adversarial scenarios.

Numerous studies have demonstrated the effectiveness of ML techniques in detecting phishing attacks across different modalities, including emails \cite{electronics12214545,Chinta_Moore_Karaka_Sakuru_Bodepudi_Maka_2025}, websites \cite{10788671,10498854}, and messages \cite{10798213,SAIDAT2024248}. These models, trained on handcrafted or learned features, have shown strong performance in distinguishing phishing from legitimate content. In the domain of voice phishing, similar efforts have emerged where researchers utilize speech-to-text conversion followed by NLP and ML classification to detect deceptive call transcripts \cite{lee2023real,10901962,10851764,10.1145/3445970.3451152,kim2021voice}. These approaches typically involve supervised classifiers such as logistic regression, decision trees, or ensemble models trained on labeled vishing datasets, achieving high accuracy in many scenarios.

However, ML models that rely on natural language inputs are known to be vulnerable to adversarial attacks. Recent research in adversarial NLP has shown that subtle manipulations—such as synonym replacement, paraphrasing, or insertion of benign-looking content—can significantly degrade classifier performance while preserving the original intent of the text \cite{10.1145/3593042,alsmadi2021adversarial}. Techniques like TextFooler, BERT-Attack, and others have revealed that NLP pipelines are susceptible to semantically similar perturbations \cite{jin2019bert}, raising concerns about the reliability of these systems in adversarial settings.

With the advent of LLMs such as GPT-4 and Gemini, the landscape of adversarial content generation has further evolved \cite{kim2024llms}. LLMs can be prompted to generate deceptive or manipulative text with high fluency and contextual coherence, making them powerful tools for crafting adversarial samples \cite{10646856,Gallagher2024}. Recent work has explored LLMs' potential in generating phishing emails, social engineering content, and even toxic or biased outputs \cite{10459698,Gallagher2024}. These studies reveal both the utility and the risks posed by LLMs when misused for malicious purposes.

Despite these developments, to the best of our knowledge, no existing work has explored the potential of commercial LLMs to conduct evasive voice phishing attacks through natural language obfuscation. In particular, there is a lack of systematic evaluation on whether LLM-generated vishing transcripts can successfully deceive trained ML classifiers. Motivated by this gap, our work investigates LLM-assisted adversarial vishing attacks by prompting commercial LLMs to transform original scam transcripts into linguistically obfuscated versions. We then assess their ability to evade detection while maintaining semantic consistency, providing a novel perspective on the threat landscape posed by modern LLMs against cybercrime.

\section{Threat Model}\label{sec:threat_model}

\begin{figure}[h]
\centering
\includegraphics[width=0.95\textwidth]{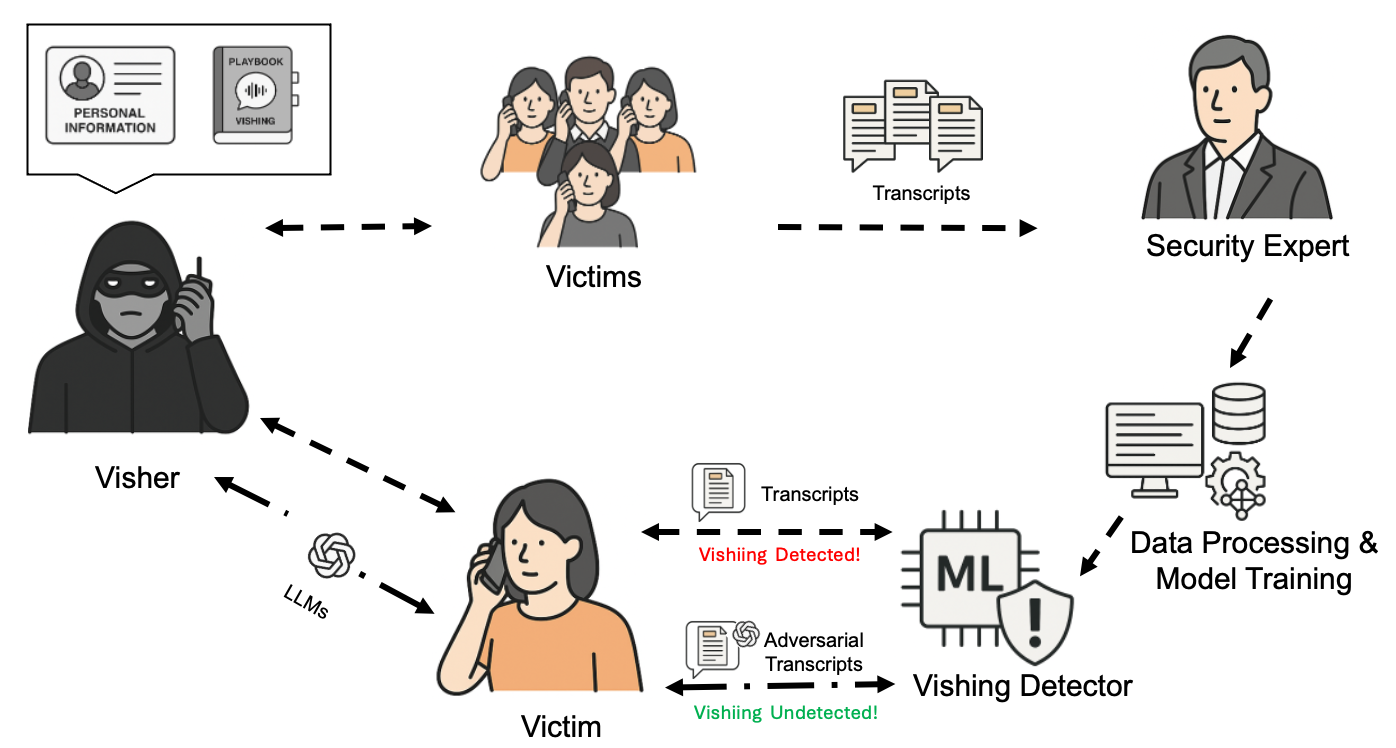}
\caption{Threat Model Overview: LLM-Generated Adversarial Vishing Transcripts Against ML-Based Classifiers.}
\label{fig:approach}
\end{figure}

In this section, we define the threat model underlying our study of LLM-assisted vishing attacks, as illustrated in Fig.~\ref{fig:approach}.

\textbf{Threat Actors. }We consider a typical vishing scenario involving a malicious actor, referred to as the \textit{visher}, who makes deceptive phone calls with the intent to extract sensitive information such as banking credentials, identity details, or authentication codes. The visher operates with a precompiled \textit{playbook}—a repository of vishing scripts on various fraudulent topics, customized to deceive different categories of victims. These scripts are informed by previously acquired personal information about the victims, such as their affiliations, transaction history, or public records.

\textbf{Data Collection by Defenders. }To defend against such threats, \textit{security experts} continuously monitor and collect transcripts from real-world vishing calls. These transcripts are derived from recorded victim interactions and processed to form labeled datasets. These datasets are then used to train ML models—referred to as \textit{vishing detectors}—that can automatically classify ongoing conversations as malicious or benign.

\textbf{Actors’ Capabilities. }In our threat model, the visher adapts to this evolving defense landscape. By leveraging powerful LLMs, the attacker refines and augments original vishing playbook scripts into \textit{adversarial transcripts}. These LLM-generated scripts are crafted to retain the deceptive intent while evading detection mechanisms by paraphrasing, reordering content, or adding benign context.

\textbf{Adversarial Dynamics. }In a conventional setting, vishing detectors deployed on the victim’s device or at the telecom backend would flag suspicious calls based on features extracted from the conversation transcript. However, when adversarial transcripts are used—crafted using LLMs to mimic legitimate communication styles while embedding malicious intent—these detectors may fail to identify the threat. As a result, the system may incorrectly classify the conversation as benign, allowing the visher to bypass security filters.

\textbf{Attacker Objectives. }The ultimate goal of the attacker is to use LLM-generated transcripts to construct highly convincing voice phishing scripts that are both contextually relevant and capable of bypassing ML-based detectors. This undermines the effectiveness of conventional detection pipelines and introduces a new class of evasive social engineering threats.

\section{Proposed Methodology}\label{sec:methodology}

This section illustrates our approach towards exploring the capabilities of commercial LLMs in deceiving the ML classifiers on voice phishing. Our approach consists of five distinct phases that systematically transform original vishing transcripts into adversarial variants while evaluating their effectiveness against trained ML classifiers and the semantic meaning preservation. The Algorithm~\ref{alg:llm_vishing_attack} provides the formal specification of our approach, which operates through five sequential phases.

\begin{algorithm}
\caption{LLM-Based Adversarial Attack on Voice Phishing Classifiers}
\scriptsize
\label{alg:llm_vishing_attack}
\begin{algorithmic}[1]
\Require Original vishing transcript $T_{orig}$
\Require LLM $\mathcal{M}$
\Require Prompt engineering strategy $\mathcal{P}$
\Require ML classifier set $\mathcal{C} = \{c_1, c_2, ..., c_n\}$
\Ensure Adversarial transcript $T_{adv}$ with evaluation metrics

\Function{AdversarialVishingAttack}{$T_{orig}, \mathcal{M}, \mathcal{P}, \mathcal{C}$}
    \State $T_{adv} \gets \emptyset$ \Comment{Adversarial transcript}
    \State $\mathcal{A} \gets \emptyset$ \Comment{Accuracy drop results}
    
    \State \textbf{Phase 1: Adversarial Prompt Construction}
    \State $P_{rephrase} \gets \Call{RephraseStrategy}{\mathcal{P}}$ \Comment{Linguistic obfuscation}
    \State $P_{noise} \gets \Call{NoiseInjection}{\mathcal{P}}$ \Comment{Benign context injection}
    \State $P_{combined} \gets P_{rephrase} \oplus P_{noise}$ \Comment{Combined prompt strategy}
    
    \State \textbf{Phase 2: LLM Generation}
    \State $T_{adv} \gets \mathcal{M}(P_{combined}, T_{orig})$ \Comment{Generate adversarial transcript}
    
    \State \textbf{Phase 3: Data Processing}
    \State $T_{adv}^{clean} \gets \Call{DataCleaning}{T_{adv}}$ \Comment{Remove noise, special chars}
    \State $tokens \gets \Call{Tokenize}{T_{adv}^{clean}}$ \Comment{MeCab tokenization}
    \State $features \gets \Call{TF-IDF}{tokens}$ \Comment{Feature vectorization}
    
    \State \textbf{Phase 4: Classifier Evaluation}
    \For{each classifier $c_i \in \mathcal{C}$}
        \State $acc_{orig}^i \gets \Call{Accuracy}{c_i, D_{original}}$ \Comment{Original accuracy}
        \State $acc_{adv}^i \gets \Call{Accuracy}{c_i, D_{adversarial}}$ \Comment{Adversarial accuracy}
        \State $acc_{drop}^i \gets acc_{orig}^i - acc_{adv}^i$ \Comment{Accuracy drop}
        \State $\mathcal{A} \gets \mathcal{A} \cup \{(c_i, acc_{drop}^i)\}$
    \EndFor
    \State $p_{wilcoxon} \gets \Call{WilcoxonSignedRankTest}{\{acc_{orig}^i\}, \{acc_{adv}^i\}}$ \Comment{Significance of attack}
    \State $R \gets \Call{RankMatrix}{\{acc_{adv}^i\}_{i=1}^{n}}$ \Comment{Compute classifier-wise accuracy ranks}
    \State $p_{friedman} \gets \Call{FriedmanTest}{R}$ \Comment{Global statistical test across LLMs}
    \If{$p_{friedman} < 0.05$}
        \State $P_{posthoc} \gets \Call{NemenyiPosthocTest}{R}$ \Comment{Pairwise significance matrix}
    \EndIf

    \State \textbf{Phase 5: Semantic Preservation Measurement}
    \State $bert_{precision} \gets \Call{BERTScore\_Precision}{T_{orig}, T_{adv}}$
    \State $bert_{recall} \gets \Call{BERTScore\_Recall}{T_{orig}, T_{adv}}$
    \State $bert_{f1} \gets \Call{BERTScore\_F1}{T_{orig}, T_{adv}}$
    \State $\mathcal{B} \gets \{bert_{precision}, bert_{recall}, bert_{f1}\}$ \Comment{BERT score metrics}
    
    \State \Return $(T_{adv}, \mathcal{B}, \mathcal{A}, p_{wilcoxon}, R, p_{friedman}, P_{posthoc})$

\EndFunction
\end{algorithmic}
\end{algorithm}

\begin{itemize}
\item \textbf{Phase 1: Adversarial Prompt Construction: }To simulate realistic LLM-assisted vishing conversations, we craft a prompt that transforms scammer speech into linguistically obfuscated and conversationally natural dialogue. We engineer sophisticated prompts that guide the LLM to perform two primary transformations: linguistic obfuscation through rephrasing ($P_{rephrase}$) and benign context injection through noise insertion ($P_{noise}$). The core transformation involves two main strategies as described in Figure \ref{fig:prompt}: rephrasing and injecting adversarial noise. Our goal in rephrasing the original scammer speech is to obscure explicit scam-related intent and make the conversation appear more legitimate. In parallel, we inject the adversarial noise, which is a contextually appropriate dialogue that expands the original speech without altering the core intent of the message. This technique serves to dilute the presence of scam-related cues by embedding them within a friendly conversation. These strategies are combined into a unified prompt ($P_{combined} = P_{rephrase} \oplus P_{noise}$) that instructs the LLM to maintain malicious intent while appearing benign to automated classifiers.

\begin{figure}[H]
\centering
\includegraphics[width=0.8\textwidth]{figures/prompt.pdf}
\caption{Vishing Generation Prompt.} \label{fig:prompt}
\end{figure}

\item \textbf{Phase 2: LLM Generation: }The constructed prompt and original transcript are processed by the target LLM ($\mathcal{M}$) to generate the adversarial transcript ($T_{adv} = \mathcal{M}(P_{combined}, T_{orig})$). This phase leverages the LLM's natural language understanding and generation capabilities to create linguistically sophisticated evasions. The generated output transforms vishing indicators into benign conversational patterns while preserving the underlying deceptive structure. We evaluate multiple state-of-the-art LLMs including GPT-4o, GPT-4o mini, Gemini 2.0, and Qwen2.5 to assess the generalizability of our approach across different model architectures and capabilities.

\item \textbf{Phase 3: Data Processing: }The data processing module is responsible for preparing raw textual input for classification by systematically cleaning and transforming it into a structured format. The generated adversarial transcripts undergo systematic preprocessing identical to the original dataset preparation used for training the ML classifiers. Given the visher's speech, this process begins with data cleaning, which involves the removal of irrelevant or redundant elements such as numbers, special characters, punctuation marks, duplicate entries, and personally identifiable information like phone numbers. Then, the text is tokenized using the MeCab-ko \cite{mecab} morphological analyzer, a tool that provides efficient processing of Korean text. Moreover, we remove common Korean stop-words with little semantic value in the context of vishing detection are removed. Following preprocessing, we apply the Term Frequency-Inverse Document Frequency (TF-IDF) technique to embed the extracted tokens. Finally, the resulting feature vectors are then passed to the classifier, which determines whether the input corresponds to a benign or malicious (scam) conversation, thus, following the methodology presented in \cite{10901962}.

\item \textbf{Phase 4: Classifier Evaluation: }Each adversarial transcript is evaluated against an ensemble of trained ML classifiers ($\mathcal{C} = \{c_1, c_2, ..., c_n\}$) that were trained on the original dataset. For each classifier $c_i$, we calculate the accuracy drop as $acc_{drop}^i = acc_{orig}^i - acc_{adv}^i$, where $acc_{orig}^i$ and $acc_{adv}^i$ denote the classifier's accuracy on the original and adversarial samples, respectively. This ensemble includes linear, tree-based, and boosting models, offering a broad view of evasion effectiveness. 

To statistically verify the effectiveness of adversarial attacks, we apply the \textit{Wilcoxon signed-rank test} between the original and adversarial accuracy distributions. A significant p-value ($p < 0.05$) indicates that the adversarial attack consistently degrades model performance across classifiers. 

To further assess whether different LLMs cause distinguishable impacts on classifier performance, we construct a rank matrix $R$ of adversarial accuracies and conduct a \textit{Friedman test}. If significant differences are detected, a \textit{Nemenyi post-hoc test} is conducted to reveal pairwise significance between LLM variants. This multi-stage evaluation not only confirms the overall attack effectiveness but also compares the relative strength of different LLM-based attack strategies.

\item \textbf{Phase 5: Semantic Preservation Measurement: }We quantify the semantic similarity between original and adversarial transcripts using comprehensive BERTScore metrics to ensure that adversarial transformations maintain the core vishing intent and contextual meaning. Specifically, we calculate three key metrics: BERTScore precision ($bert_{precision}$), BERTScore recall ($bert_{recall}$), and BERTScore F1 ($bert_{f1}$), which together form our semantic preservation measurement set ($\mathcal{B} = \{bert_{precision}, bert_{recall}, bert_{f1}\}$). These metrics provide a comprehensive assessment of semantic preservation quality by comparing contextualized embeddings of the original and generated texts.

\end{itemize}

Finally, we provide a detailed case study on the Analysis of Original vs. Adversarial Transcripts to demonstrate the methodology for examining semantic preservation and evasion strategies employed by our LLM-based approach. This analysis involves a comparison of original and adversarial transcripts, examining how strategic rephrasing and benign context injection are implemented and analyzing the effectiveness of proposed method accordingly.

\section{Experimental Setup}\label{sec:exp}
This section outlines the experimental setup of our study, including the dataset used, the LLMs employed, the evaluation metrics applied, and the ML classifiers for vishing detection, along with their performance on the original transcripts.

\subsection{Dataset}
\label{subsec:dataset}
In this study, we utilize a balanced subset of the KorCCViD v1.3 dataset \cite{KorcVV}, consisting of 609 transcripts from vishing scenarios and 609 from non-vishing scenarios. The vishing samples are derived from real-world Korean scam call transcripts, while the benign samples represent typical everyday conversational speech. This dataset captures realistic vishing contexts and offers a robust foundation for evaluating semantic-preserving adversarial attacks. The data is randomly partitioned into training, validation, and testing sets, with 779 samples allocated for training, 195 for validation, and 244 for testing.

\subsection{Used LLMs}
\label{subsec:llms}
We evaluated our attack using 4 different LLMs that follows our defined prompt as presented in Fig.~\ref{fig:prompt}: GPT4-o and GPT4-o mini \cite{openai2023gpt4}, Gemini 2.0 \cite{google2023gemini}, and Qwen2.5 \cite{bai2023qwen}. We selected our LLMs based on several key factors, including model size, architecture, and language abilities. These models represent a range of capabilities and have been widely used in previous research.

\subsection{Evaluation Metrics}\label{sec:metrics}
To evaluate the effectiveness of LLM-generated adversarial transcripts and ensure semantic fidelity with the original vishing content, we adopt three sets of metrics:

\textbf{1. Classifier Performance Metrics.} These metrics quantify how adversarial transcripts impact vishing detection models:

\begin{itemize}
    \item \textbf{Standard Classification Metrics:} We compute precision, recall, accuracy, and F1-score on both original and adversarial datasets:

    \begin{equation}
    \text{Precision} = \frac{TP}{TP + FP},  \quad \text{Recall} = \frac{TP}{TP + FN}
    \end{equation}

    \begin{equation}
    \text{Accuracy} = \frac{TP + TN}{TP + TN + FP + FN}
    \end{equation}

    \begin{equation}
    F1 = 2 \cdot \frac{\text{Precision} \cdot \text{Recall}}{\text{Precision} + \text{Recall}}
    \end{equation}

    \item \textbf{Accuracy Drop ($\Delta \text{Accuracy}$):} Measures the performance degradation of classifiers caused by adversarial transcripts. For each classifier $c_i$:

    \begin{equation}
    \Delta \text{Accuracy}_i = \text{Accuracy}_{\text{original}}^i - \text{Accuracy}_{\text{adversarial}}^i
    \end{equation}
\end{itemize}

\textbf{2. Statistical Testing Metrics.}  
To assess whether the classification performance degradation across different LLM-generated adversarial transcripts is statistically significant, we employ non-parametric statistical testing. These methods evaluate the consistency and strength of adversarial impact across all classifiers:

\begin{itemize}
    \item \textbf{Wilcoxon Signed-Rank Test:} To evaluate the effectiveness of each individual LLM attack, we perform a one-tailed Wilcoxon signed-rank test comparing the original and adversarial accuracies across all classifiers. This non-parametric test assesses whether adversarial examples consistently lead to a reduction in classifier performance. The test statistic is defined as:
    \begin{equation}
    W = \min\left(W_+, W_-\right)
    \end{equation}
    where \( W_+ \) and \( W_- \) are the sums of ranks for positive and negative accuracy differences, respectively. Since our hypothesis is directional (i.e., adversarial accuracy is expected to be lower than the original), a one-tailed $p$-value is computed as:
    \begin{equation}
    p = \mathbb{P}(W \leq w)
    \end{equation}
    A small $p$-value (e.g., $p < 0.05$) indicates that the adversarial attack produces a statistically significant and consistent drop in classifier accuracy.

    \item \textbf{Friedman Test:} This non-parametric test is used to determine whether there are overall significant differences in classifier accuracy under different LLM attacks. Given $k$ attack models and $n$ classifiers, we first compute ranks $R_{i,j}$ of the adversarial accuracies for each classifier $i$ across $k$ LLMs (lower accuracy implies a stronger attack and thus a higher rank). The Friedman test statistic is calculated as:

    \begin{equation}
    \chi_F^2 = \frac{12n}{k(k+1)} \left[\sum_{j=1}^{k} \bar{R}_j^2\right] - 3n(k+1)
    \end{equation}

    where $\bar{R}_j$ denotes the average rank of LLM $j$.

    \item \textbf{Average Ranks:} The mean rank for each LLM model is computed to indicate its relative adversarial strength. A lower average rank indicates stronger attack efficacy:

    \begin{equation}
    \bar{R}_j = \frac{1}{n} \sum_{i=1}^{n} R_{i,j}
    \end{equation}

    These ranks are used as the basis for pairwise comparison in the next step.

    \item \textbf{Nemenyi Post-hoc Test:} If the Friedman test reveals significant overall differences, we perform the Nemenyi test to compare each pair of LLMs. The test returns a matrix of adjusted p-values, where each entry indicates the statistical significance of performance difference between two LLM attacks:

    \begin{equation}
    p_{j_1, j_2} = \text{P-value comparing } \bar{R}_{j_1} \text{ and } \bar{R}_{j_2}
    \end{equation}
\end{itemize}

\textbf{3. Semantic Similarity Metrics.} To ensure that adversarial texts preserve the core malicious intent and semantics of the originals, we apply BERTScore:

\begin{itemize}
    \item Given an original transcript $T_{\text{orig}} = [r_1, ..., r_m]$ and adversarial transcript $T_{\text{adv}} = [c_1, ..., c_n]$, contextual embeddings $\vec{r}_i$ and $\vec{c}_j$ are obtained using a pre-trained BERT model. Cosine similarity is calculated as:

    \begin{equation}
    \text{sim}(\vec{r}_i, \vec{c}_j) = \frac{\vec{r}_i \cdot \vec{c}_j}{\|\vec{r}_i\| \|\vec{c}_j\|}
    \end{equation}

    \item \textbf{BERTScore Precision, Recall, and F1} are defined as:

    \begin{equation}
    \text{BERTScore}_{\text{Precision}} = \frac{1}{n} \sum_{j=1}^{n} \max_i \text{sim}(\vec{r}_i, \vec{c}_j)
    \end{equation}

    \begin{equation}
    \text{BERTScore}_{\text{Recall}} = \frac{1}{m} \sum_{i=1}^{m} \max_j \text{sim}(\vec{r}_i, \vec{c}_j)
    \end{equation}

    \begin{equation}
    \text{BERTScore}_{F1} = 2 \cdot \frac{\text{BERTScore}_{\text{Precision}} \cdot \text{BERTScore}_{\text{Recall}}}{\text{BERTScore}_{\text{Precision}} + \text{BERTScore}_{\text{Recall}}}
    \end{equation}
\end{itemize}

\subsection{ML Classifiers }
\label{subsec:classifiers}
We trained several ML classifiers on the proposed dataset using a consistent data split configuration, incorporating both linear and ensemble-based models to enable a comprehensive evaluation. Table~\ref{tab:mlperfor} presents their performance on the test set across four metrics: F1-score, precision, recall, and accuracy.

The results indicate consistently high performance across all models, with test accuracies ranging from approximately 95\% to 99.6\%. This strong performance can be attributed to two primary factors: (1) the KorCCViD v1.3 dataset is perfectly balanced across classes, which helps mitigate classification bias; and (2) the models showed no signs of overfitting, as demonstrated by their robust generalization to the unseen test set.

\begin{table}[H]
\centering
\caption{Performance  of various ML classifiers}
\begin{tabular}{lcccc}
\toprule
\textbf{} & \textbf{F1 Score} & \textbf{Precision} & \textbf{Recall} & \textbf{Accuracy} \\
\midrule
LogisticRegression & 0.991935 & 0.991803 & 0.991803 & 0.991803 \\
DecisionTree & 0.951305 & 0.950820 & 0.950806 & 0.950820 \\
RandomForest & 0.988000 & 0.987705 & 0.987703 & 0.987705 \\
AdaBoost & 0.983607 & 0.983607 & 0.983607 & 0.983607 \\
GradientBoosting & 0.955683 & 0.954918 & 0.954899 & 0.954918 \\
HistGradientBoosting & 0.979540 & 0.979508 & 0.979508 & 0.979508 \\
XGB & 0.979540 & 0.979508 & 0.979508 & 0.979508 \\
LGBM & 0.983737 & 0.983607 & 0.983605 & 0.983607 \\
CatBoost & 0.959510 & 0.959016 & 0.959005 & 0.959016 \\
LinearSVC & 0.995935 & 0.995902 & 0.995902 & 0.995902 \\
\bottomrule
\end{tabular}
\label{tab:mlperfor}
\end{table}

We primarily use classical ML classifiers due to their widespread practical adoption, interpretability, and low computational cost. Our goal is to show that even these lightweight models despite high baseline accuracy remain vulnerable to LLM-generated adversarial attacks. This highlights that such threats persist even in real-world, resource-efficient deployments, with broader implications for both traditional and modern NLP-based defenses.

\section{Results}\label{sec:results}
In this section, we present a detailed analysis of the proposed approach, including a comparison of performances of ML classifiers trained on the original and LLM-generated vishing transcripts with statistical testing as well as the semantic similarity of adversarial transcripts. In addition, we provide a case study with original and adversarial transcripts and the costs for conducting such attacks.

\subsection{Adversarial Effectiveness and Semantic Similarity}

Table~\ref{tab:llm_perf} shows the classification accuracy of various models on 100 adversarial vishing transcripts generated by four different LLMs. To evaluate the impact of each model, we calculate the average accuracy drop across ten classifiers. As shown in the last rows of Table~\ref{tab:llm_perf}, Qwen2.5 results in the highest average accuracy drop at 33.83\%, indicating its strong evasion capability. GPT-4o follows with a 16.16\% drop, while Gemini 2.0 and MiniGPT-4o yield more moderate drops of 7.18\% and 3.42\%, respectively.

To evaluate whether each LLM-generated adversarial attack leads to a statistically significant reduction in classifier performance, we conduct one-tailed Wilcoxon signed-rank tests comparing the original and adversarial accuracies across all classifiers. As shown in Table~\ref{tab:llm_perf}, all four LLMs demonstrate statistically significant performance degradation, with one-tailed $p$-values below the 0.05 threshold. In particular, GPT-4o, Gemini 2.0, and Qwen2.5 yield highly significant reductions with $p = 0.0010$, while MiniGPT-4o also achieves significance with $p = 0.0098$. These results confirm that the observed accuracy drops are not only substantial in magnitude but also statistically consistent across classifiers, validating the effectiveness of the adversarial attacks.

To assess whether these performance differences are statistically significant, we conduct a non-parametric Friedman test on classifier-wise adversarial accuracies across the four LLMs. The result yields a Friedman statistic of 28.0408 with a $p$-value of 0.000004, indicating significant differences in classifier performance under different LLM attacks. Based on per-row (classifier-wise) rankings of adversarial effectiveness, Qwen2.5 achieves the lowest average rank (1.0), followed by GPT-4o (2.0), Gemini 2.0 (3.3), and MiniGPT-4o (3.7). 

\begin{figure}[h]
\centering
\includegraphics[width=0.65\linewidth]{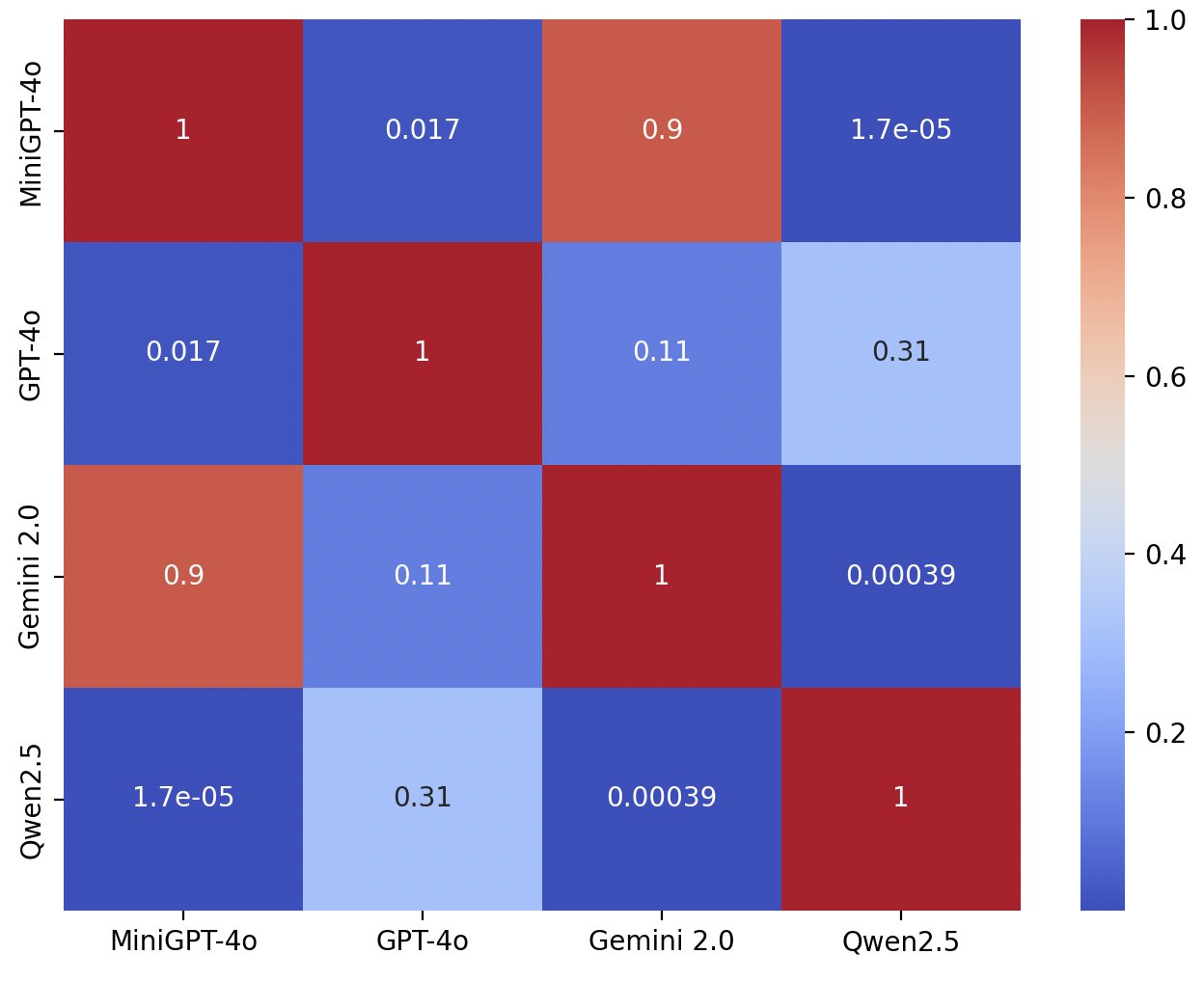}
\caption{
Nemenyi post-hoc test results comparing adversarial effectiveness of four LLMs based on classifier accuracy rankings. Each cell displays the $p$-value of the pairwise comparison between two LLMs. Statistically significant differences ($p < 0.05$) are observed between Qwen2.5 and all other models, as well as between GPT-4o and MiniGPT-4o. Darker blue regions indicate stronger statistical significance.
}
\label{fig:nemenyi_heatmap}
\end{figure}

To further identify pairwise differences, we perform a Nemenyi post-hoc test. As illustrated in Figure~\ref{fig:nemenyi_heatmap}, Qwen2.5's attack performance is significantly stronger than that of MiniGPT-4o ($p < 0.001$) and Gemini 2.0 ($p < 0.001$). GPT-4o is also significantly more effective than MiniGPT-4o ($p = 0.017$), while its difference from Qwen2.5 and Gemini 2.0 is not statistically significant. These findings confirm that Qwen2.5 is the most disruptive adversarial generator, with GPT-4o as the second most effective.

\begin{figure}[H]
\includegraphics[width=\textwidth]{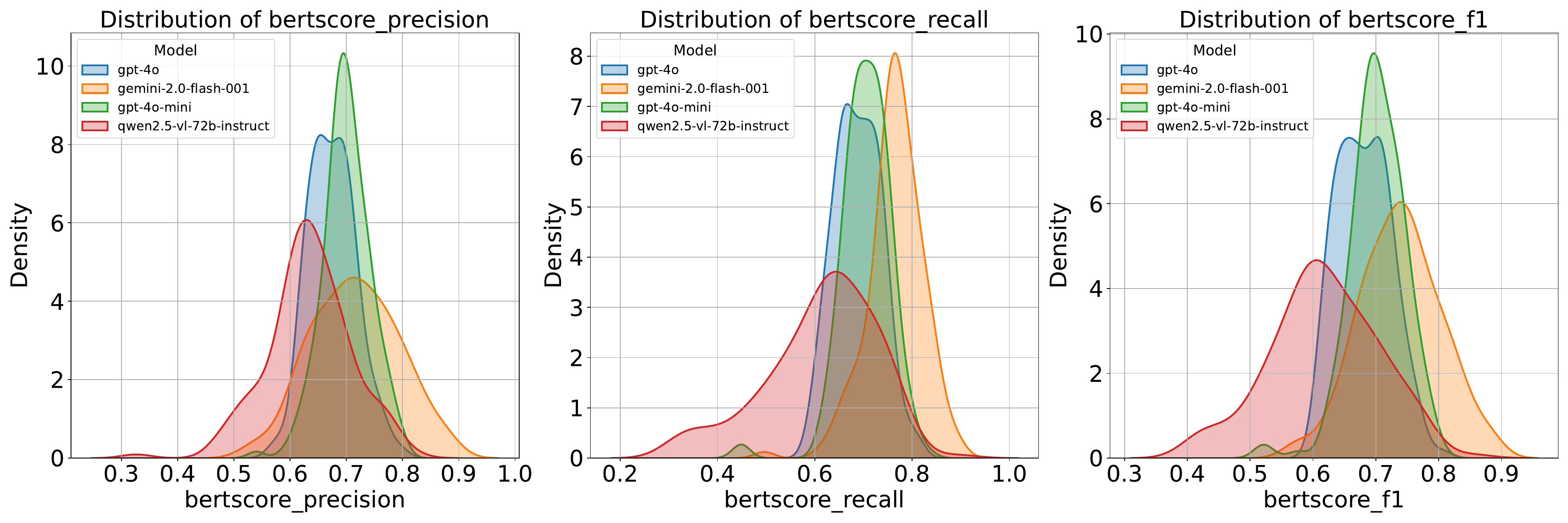}
\caption{Bert Score Between Original Transcripts and LLM perturbed ones.} \label{fig:bertscore_distribution}
\end{figure}

While Qwen2.5 demonstrates the strongest evasion performance, its semantic fidelity is relatively poor. As illustrated in Figure~\ref{fig:bertscore_distribution}, Qwen2.5-generated texts exhibit a wide range of BERTScore values (from 0.45 to 0.85, peaking around 0.65), indicating frequent deviations from the original transcript’s meaning. In many instances, it introduces off-topic or incoherent content that disrupts the intended prompt structure and alters the vishing context.

In contrast, GPT-4o achieves a strong balance between adversarial effectiveness and semantic preservation. Although it causes the second-highest accuracy drop, it maintains high BERTScore precision, recall, and F1 values (ranging from 0.72 to 0.75). This suggests that GPT-4o-generated transcripts successfully preserve the core malicious intent while introducing meaningful adversarial variations.

Given this trade-off, we select GPT-4o as the representative LLM for subsequent evaluations. It demonstrates statistically validated evasion capability without compromising semantic integrity—an essential criterion for generating high-quality adversarial examples in vishing scenarios.

\begin{table}[H]
\centering

\caption{Performance comparison of various classifiers on 100 vishing samples}
\begin{tabular}{l|c|cccc}
\toprule
\textbf{Classifier} & \multirow{2}{*}{\textbf{Original Acc.}} & \multicolumn{4}{c}{\textbf{Adversarial Acc.}} \\
\cmidrule(lr){3-6}
& & \textbf{MiniGPT-4o} & \textbf{GPT-4o} & \textbf{Gemini 2.0} & \textbf{Qwen2.5} \\
\midrule
LogisticRegression     & 0.991803 & 0.958904 & 0.760274 & 0.773973 & 0.623288 \\
DecisionTree           & 0.950820 & 0.890411 & 0.726027 & 0.856164 & 0.458904 \\
RandomForest           & 0.987705 & 0.986301 & 0.979452 & 0.986301 & 0.732877 \\
AdaBoost               & 0.983607 & 0.945205 & 0.883562 & 0.938356 & 0.630137 \\
GradientBoosting       & 0.954918 & 0.815068 & 0.623288 & 0.842466 & 0.445205 \\
HistGradientBoosting   & 0.979508 & 0.986301 & 0.849315 & 0.958904 & 0.801370 \\
XGB                    & 0.979508 & 0.952055 & 0.876712 & 0.952055 & 0.746575 \\
LGBM                   & 0.983607 & 0.986301 & 0.808219 & 0.965753 & 0.726027 \\
CatBoost               & 0.959016 & 0.945205 & 0.856164 & 0.958904 & 0.561644 \\
LinearSVC              & 0.995902 & 0.958904 & 0.787671 & 0.815068 & 0.657534 \\
\midrule
\textbf{Average Acc. Drop} & -- & 3.42\% $\downarrow$ & 16.16\% $\downarrow$ & 7.18\% $\downarrow$ & 33.83\% $\downarrow$ \\
\midrule
\textbf{Wilcoxon p-value} & -- & 0.0098 & 0.0010 & 0.0010 & 0.0010 \\
\midrule
\textbf{Average Ranks} & -- & 3.7 & 2.0 & 3.3 & 1.0 \\
\bottomrule
\end{tabular}
\label{tab:llm_perf}
\end{table}

\subsection{Full Vishing Dataset Evaluation Using GPT-4o}
We evaluated our adversarial attack on the full set of vishing transcripts to assess the effectiveness of GPT-4o in deceiving ML classifiers. As shown in Table~\ref{tab:llm_gpt4o_perf}, the average classification accuracy across all models dropped from 97.66\% to 81.35\%, reflecting a substantial degradation. 

Individual model performance on adversarial vishing samples ranged from 64.53\% (GradientBoostingClassifier) to 95.89\% (RandomForestClassifier), corresponding to accuracy drops between 2.88\% and 30.96\%. This indicates that GPT-4o was successful in crafting semantically consistent adversarial transcripts that caused a measurable decline in classifier reliability. To statistically validate the effectiveness of GPT-4o-generated adversarial examples on full dataset, we also performed a one-tailed Wilcoxon signed-rank test comparing original and adversarial accuracies. The test confirmed a consistent performance drop, yielding a significant $p$-value of $0.00098$ ($p < 0.05$).

\begin{figure}[H]
    \centering
    \captionsetup[subfloat]{labelformat=empty}

    \begin{minipage}[b]{0.48\textwidth}
        \centering
        \subfloat[Original Vishing Transcripts]{\includegraphics[width=\textwidth]{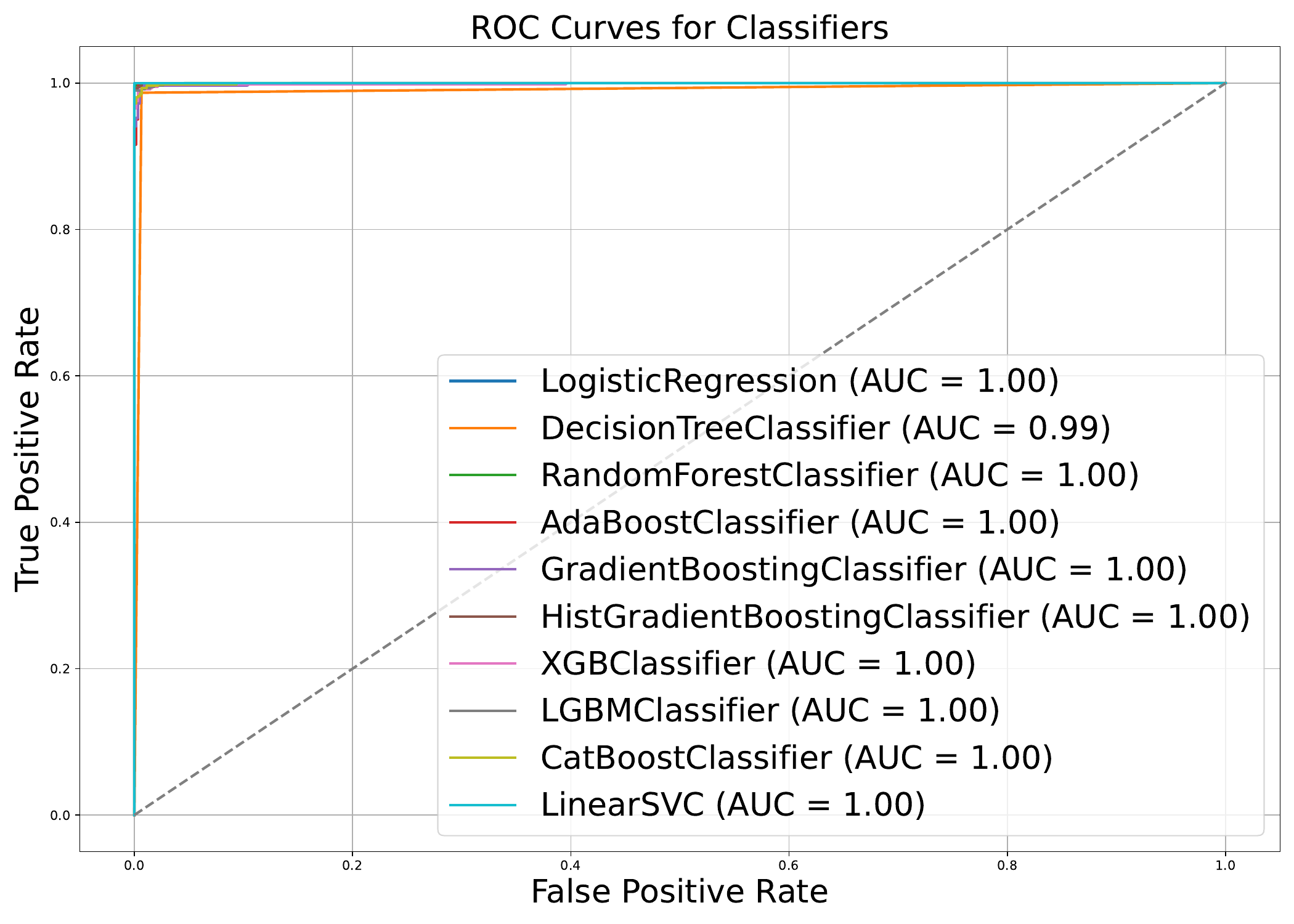}}
    \end{minipage}
    \hfill
    \begin{minipage}[b]{0.48\textwidth}
        \centering
        \subfloat[Adversarial Vishing Transcripts]{\includegraphics[width=\textwidth]{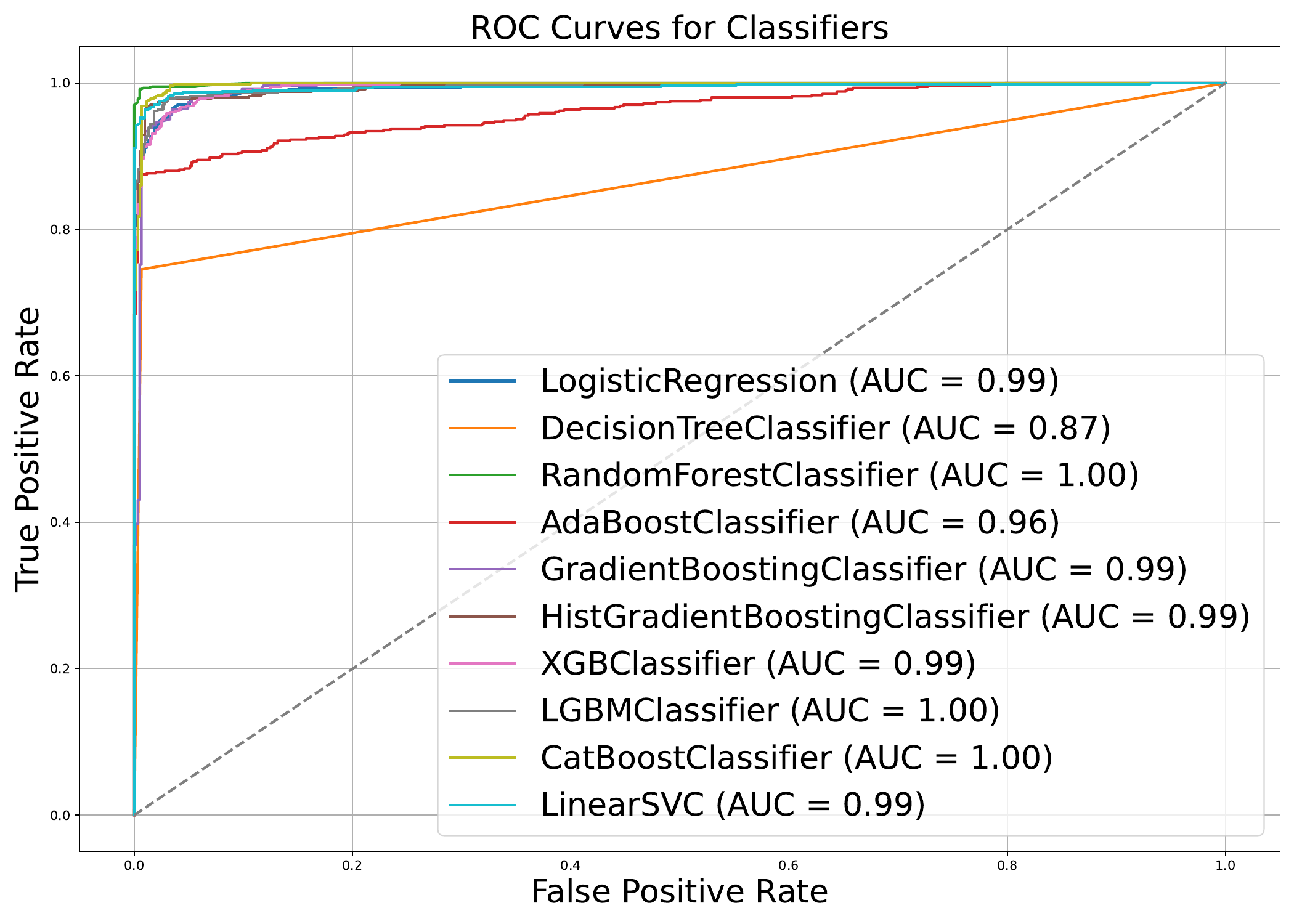}}
    \end{minipage}
    \caption{ROC curves for original and adversarial vishing transcripts.}
    \label{fig:roc}
\end{figure}

To further investigate the impact of adversarial perturbations, we examined ROC curves before and after applying GPT-4o-based obfuscation, as illustrated in Fig.~\ref{fig:roc}. Notably, the AUC values of DecisionTreeClassifier and AdaBoostClassifier declined to 0.87 and 0.96, respectively. This suggests that several adversarial vishing scenarios were misclassified as benign, increasing the false negative rate and reducing overall detection performance.

\begin{table}[H]
\centering
\caption{Performance comparison of classifiers on original vs. GPT-4o adversarial vishing samples}
\begin{tabular}{l|ccc}
\toprule
\textbf{Classifier} & \textbf{Original Acc.} & \textbf{Adversarial Acc.} & \textbf{Acc. Drop} \\
\midrule
LogisticRegression                & 0.991803 & 0.763547 & 0.228256 \\
DecisionTreeClassifier            & 0.950820 & 0.745484 & 0.205336 \\
RandomForestClassifier            & 0.987705 & 0.958949 & 0.028756 \\
AdaBoostClassifier                & 0.983607 & 0.834154 & 0.149453 \\
GradientBoostingClassifier        & 0.954918 & 0.645320 & 0.309598 \\
HistGradientBoostingClassifier    & 0.979508 & 0.857143 & 0.122365 \\
XGBClassifier                     & 0.979508 & 0.844007 & 0.135501 \\
LGBMClassifier                    & 0.983607 & 0.862069 & 0.121538 \\
CatBoostClassifier                & 0.959016 & 0.844007 & 0.115009 \\
LinearSVC                         & 0.995902 & 0.779967 & 0.215935 \\
\bottomrule
\end{tabular}
\label{tab:llm_gpt4o_perf}
\end{table}

\subsection{Analysis of Original vs. Adversarial Transcripts}
We validate our given prompt to the LLM to show that the adversarial text generated by GPT-4o preserves the meaning while effectively fools the ML classifiers. We present in Table~\ref{tab:comp} one of the generated adversarial vishing transcripts. For easier understanding, we have included the English translation of both Korean texts. Using different color labels, we further clarify every aspect of the text modification done by the LLMs, such as paraphrasing highly vishing terms indicators using the red color, as well as presenting the benign added sentences in green. One of the key pieces of evidence of the context preservation in this shown example is that terms like banks, account, and business license are preserved or paraphrased. Another example is the following transformation: \textit{“Kookmin Bank involving card and securities concerns”} $\longrightarrow$ \textit{“Kookmin Bank had similar cases, and multiple people shared concerns”}. Although the words change, the core message stays intact. On the other side, we show in Table~\ref{tab:comp} the benign marked tokens in green, such as \textit{“feel free to ask”}, \textit{“I’ve been looking into it”}, are been added to make the transcript more friendly and neutral. In addition, compared to prior works that could not explain the reason behind why LLM success in their given task, we demonstrated through Table~\ref{tab:comp} that our generated transcript preserves meaning while deceiving the classifier.

By exploiting the characteristics of the encoding technique, the generated transcripts successfully evaded detection by the classifier. Since our classification model relies on token occurrence patterns, uniqueness, and the overall length of the transcript as emphasized through TF-IDF, rephrasing and injecting additional bengin statements altered these statistical features, thus evading the detection.

In addition, from practical perspective, we evaluated the resource requirements for executing GPT-4o-based adversarial attacks. The average cost to generate a single adversarial transcript using GPT-4o was approximately \$0.00685, with an average generation time of 8.595 seconds. These figures highlight the economic feasibility and scalability of such attacks. An adversary with limited financial and computational resources could feasibly launch large-scale evasive vishing campaigns by leveraging commercial LLMs as attack enablers, making this threat vector particularly concerning in practice.

Furthermore, our empirical observations revealed that all tested commercial LLMs—including GPT-4o, GPT-4o-mini, Gemini 2.0, and Qwen2.5—responded to our adversarial prompts without issuing rejections or security-related warnings. Despite the adversarial intent embedded in the prompts, none of the models triggered content filters or exhibited refusal behaviors. This raises critical concerns regarding the effectiveness of existing safety guardrails in current LLM deployments

\begin{table}[H]
\centering
\caption{Analysis of Original and Adversarial Text Samples, \textcolor{rephrased}{Red} means the rephrases parts, \textcolor{neutral}{Blue} means the neutral tone, and \textcolor{added}{Green} means the added benign words. }
\begin{adjustbox}{max width=\textwidth}
\begin{tabular}{|>{\raggedright\arraybackslash}p{3cm}|>{\raggedright\arraybackslash}p{5cm}|>{\raggedright\arraybackslash}p{4cm}|>{\raggedright\arraybackslash}p{5cm}|}
\hline
\multicolumn{2}{|c|}{\textbf{Original Text}} & \multicolumn{2}{c|}{\textbf{Adversarial Text}} \\
\hline
\textbf{Korean} & \textbf{English Translation} & \textbf{Korean} & \textbf{English Translation} \\
\hline
\begin{CJK}{UTF8}{mj}
\textcolor{rephrased}{농협 하나 통장} 여기 피해 자분 확인 통장 대해서 굉장히 거래처 방문 고서 통장 여보 많이 으시 에서 결정 공부 다고 습니다 일단 \textcolor{rephrased}{국민은행} 에서 으로 얘기 \textcolor{rephrased}{카드 증권} 아무래도 사건 단체 에서 직원 많이 열받 으시 \textcolor{rephrased}{사업자 등록증} 면서 다른 친구 \textcolor{rephrased}{인터넷 뱅킹} 으로 지속 으로 사건 라고
\end{CJK}
&
\textcolor{rephrased}{NongHyup and Hana Bank accounts} were involved. The victim was identified, and the bank account information was verified. There were heavy interactions with clients. The conversation reportedly came from \textcolor{rephrased}{Kookmin Bank} involving \textcolor{rephrased}{card and securities} concerns. Due to the nature of the incident, many staff were furious. A \textcolor{rephrased}{business license} was shown, and another friend had been involved through \textcolor{rephrased}{online banking}, suggesting the fraud continued through multiple fronts.
&
\begin{CJK}{UTF8}{mj}
\textcolor{rephrased}{농협 관련 통장} 어서요 요즘 \textcolor{neutral}{통장 어떻게 사용} 조금 \textcolor{neutral}{궁금} 상태 여기저기 제일 많이 쓰이 \textcolor{neutral}{통장 하나 아서요} 거래처 \textcolor{neutral}{통장 자주 사용} 어떤 경우 약간 \textcolor{neutral}{문제 생긴} 거든요 그때 \textcolor{rephrased}{친구 피해} 아서 어떻게 해결 \textcolor{added}{생각 많이 더랬어요} 그리고 \textcolor{rephrased}{국민은행 요즘 비슷 이야기} 여러 사람 경우 \textcolor{neutral}{불안해하 친구 얘기} 나누 면서 \textcolor{neutral}{문제} 라고 마다 조금 다르 지만 \textcolor{neutral}{관련 어서 조심 분위기} 네요 \textcolor{rephrased}{사업자 등록증 얘기} 나왔 혹시 으신 \textcolor{rephrased}{인터넷 뱅킹 관련} 돼서 \textcolor{neutral}{궁금 여쭤 려고} 그런 자주 \textcolor{neutral}{문제} 그래도 사람 마다 \textcolor{added}{생각 다를 어서요} 아무래도 부분 대해서 서로 빠르 의견 주고받 으면서 해결 방법 으면 혹시 필요 정보 다면 \textcolor{added}{편하 세요 요즘 시간 어서 알아보 려고}
\end{CJK}
&
\textcolor{rephrased}{This is about a NongHyup-related account}. Lately, I'm \textcolor{neutral}{curious about how the account is being used}. It’s one of the most frequently used \textcolor{neutral}{bank accounts}, especially for business transactions. Some \textcolor{neutral}{issues occurred} when a \textcolor{rephrased}{friend was affected}, and we \textcolor{added}{thought about how to resolve it}. Also, \textcolor{rephrased}{Kookmin Bank has had similar cases}, and multiple people \textcolor{neutral}{shared concerns}. When we talked about it with a \textcolor{neutral}{friend}, it was clear that each case is different, but \textcolor{neutral}{there's a general sense of caution}. Someone even mentioned a \textcolor{rephrased}{business registration certificate}, and I wanted to ask about \textcolor{rephrased}{online banking issues}. These problems happen often. Still, \textcolor{added}{everyone has different views}, so I hope we can exchange ideas and find a solution. If you need information, \textcolor{added}{feel free to ask—I've been looking into it recently}.
\\ \hline
\end{tabular}
\end{adjustbox}
\label{tab:comp}
\end{table}

\section{Conclusion}\label{sec:con}
This study highlights the emerging threat posed by LLMs in generating evasive vishing transcripts. By prompting commercial LLMs with real-world scam scripts, we show that these models can produce linguistically obfuscated yet semantically consistent transcripts capable of bypassing state-of-the-art ML-based vishing detectors. Our evaluation reveals that such attacks are not only effective but also economically and computationally inexpensive, making them accessible to a wide range of adversaries. These findings call for the development of more robust vishing detection systems and emphasize the need for commercial LLM providers to implement safeguards that prevent prompt misuse for such malicious purposes.
%
%
%
%

\bibliographystyle{splncs04}
\bibliography{mybibliography}

\end{document}